# Acquisition and Analysis of Scanning Tunneling Spectroscopy Data – WSe₂ Monolayer

Randall M. Feenstra,[1*] Grayson R. Frazier,[1] Yi Pan,[2,3] Stefan Fölsch,[2] Yu-Chuan Lin,[4] Bhakti Jariwala,[4] Kehao Zhang,[4] and Joshua A. Robinson[4]

[1]Dept. Physics, Carnegie Mellon University, Pittsburgh, PA, 15213 U.S.A.

[2]Paul-Drude-Institut für Festkörperelektronik, Hausvogteiplatz 5-7, 10117 Berlin, Germany

[3]Center for Spintronics and Quantum Systems, State Key Laboratory for Mechanical Behavior of Materials, Xi'an Jiaotong University, Xi'an 710049, China

[4]Dept. Materials Science and Engineering, and Center for 2-Dimensional and Layered Materials, The Pennsylvania State University, University Park, PA, 16802 U.S.A.

## Abstract

Acquisition and analysis is described for scanning tunneling spectroscopy data acquired from a monolayer of WSe₂ grown on epitaxial graphene on SiC. Curve fitting of the data is performed, in order to deduce band-edge energies. In addition to describing the details of the theoretical curves used for the fitting, the acquisition and analysis methods are also discussed within the larger context of the historical development of scanning tunneling spectroscopy techniques.

## I. Introduction

Since the development of the scanning tunneling microscope (STM) in 1982 [1], spectroscopy measurements with the instrument, i.e. scanning tunneling spectroscopy (STS), have played a very important role in its application [2,3,4,5]. In the present article, for this special edition of the Journal of Vacuum Science and Technology A, we present a brief review of the development of acquisition and analysis methods for STS, particularly as it relates to measurements on semiconducting materials. We present an example of such work for recent measurements made on a monolayer of WSe₂, deposited on epitaxial graphene (EG) which was formed on SiC(0001).

Considering semiconductors with band gaps of size ~1 eV, clear delineation of the band edges (as well as any states that might exist within the band gap, e.g. due to surface and/or defect states) necessitates high dynamic range STS measurements over a 2 – 3 V range in both positive and negative sample bias voltages [6,7]. A dynamic range in the tunneling current and conductance of at least 3 – 4 orders of magnitude is required, and this must be achieved while acquiring a spectrum in a typical time interval of a few seconds (also while maintaining currents of ≲ 1 nA, to avoid sample or tip damage and/or other non-equilibrium effects). A good current preamplifier in the STM apparatus is important for this purpose, but in addition, varying the tip-sample separation by typically 1 Å/V during the measurement (moving the tip closer to the surface as the magnitude of the voltage is reduced) enables an enhancement of the dynamic range by 1 – 2 orders of magnitude [6]. Subsequent "normalization" of the spectra is then important, to remove (as much as possible)

---

this variation in tip-sample separation, thus converting the raw data into a more easily understandable spectrum that is normalized to constant tip-sample separation [7,8].

The STS spectra presented in this work were acquired from monolayer films of $WSe_2$, deposited by chemical vapor deposition on epitaxial graphene that, itself, was formed on SiC. Details of the film growth have been provided in two recent works [9,10], and will not be repeated here. The motivation of these studies can be viewed in the context of the large world-wide research effort over the past decade or so that has focused on such 2-dimensional (2D) materials. Since the pioneering work of Novoselov et al. [11], there has been great interest in such materials (including stacking monolayers of different materials one atop the other to form a "heterostructure") [12], both from a fundamental point of view and in terms of potential applications. Much of the world-wide activity has utilized flakes of material, exfoliated from bulk crystals (or in thin film form, removed from a substrate) and then transferred on to some carrier substrate. However, for large-area materials, as would be most useful for applications, growth methods such as chemical vapor deposition are required. Hence, the materials studied here were grown as part of a program devoted to low-power tunneling field-effect transistors (TFETs) made from atomically thin layers of transition metal dichalcogenide (TMD) materials [13].

STM and STS studies have played an important role in the study of monolayer (or few-layer) TMD materials. Historically, even in the very early years of the STM, studies of bulk TMD crystals were performed [14,15]. The fact that the surfaces of such materials as relatively nonreactive (i.e. at least compared to typical semiconducting materials such as Si or GaAs) meant that a clean surface could be prepared by removing (exfoliating) some uppermost layers from the crystal, and then rapidly transferring the sample into vacuum (or, in some cases, just studying it in ambient air). More recently, STM/STS work on TMD materials have shifted towards study of mono- and few-layer materials. A number of papers have reported STS results of $WSe_2$, the material studied in the present work [9,10,16,17,18,19,20]. One notable recent paper that we discuss in particular is the STS study of Hill et al. [21], performed at room temperature on monolayers of $MoS_2$ and $WS_2$ (as well as stacked bilayers of one material on the other). These authors presented a method for curve fitting their data, thus enabling determination of band-edge energies, band gaps, and band offsets between the $MoS_2$ and $WS_2$ layers. Good fits of the theoretical model to the STS data were obtained.

In the present work, STS data acquired from $WSe_2$ is presented, and curve fitting of the data is performed. In contrast to the prior work of Hill et al. [21], the present data was acquired at a lower temperature of 5 K (as was the data of Ref. [16]), and with much higher dynamic range. We find that these changes necessitate development of a new theory for simulating the conductance spectra. That is to say, the prior method of Hill et al., while quite adequate for their room-temperature data, is found to work poorly when applied to our low-temperature data. Certain features associated with the various electronic bands are more distinctly seen in our data. In order to fit these features, we find it necessary to substantially revise the model of Hill et al. A theory is developed in which the varying wave-vector of the electronic states is explicitly included, yielding a significant improvement in the quality of the fits. Nevertheless, a certain feature in our data remain poorly explained by this revised model, and a discussion is provided of the possible origin of this discrepancy.

## II. STS Acquisition and Analysis
## A. Acquisition and Normalization

In this Section, we provide an overview of STS acquisition and analysis methods, as relevant particularly for semiconductor materials in which high dynamic range measurements over a relative large voltage range are needed in order to clearly define the band edge features (on either side of the band gap) as well as other features that may lie within the valence band (VB) or conduction band (CB) themselves, or within the band gap. We illustrate the methods by discussion of data acquired from a monolayer of $WSe_2$ [9,10].

Early-on in STS measurements of semiconductors, it was realized that measurements at various values of tip-sample separation permit an enhancement in the dynamic range of the data [6,7,8]. Since the current varies, to lowest order, exponentially with separation according to $\exp(-2\kappa s)$ where $s$ is the separation and $\kappa$ is a constant. Typically, for vacuum tunneling with low sample bias voltages, $\kappa$ will have a value of approximately $\sqrt{2m_0\bar{\phi}}/\hbar \approx 1.1$ Å$^{-1}$, where $m_0$ is the free electron mass, $\bar{\phi}$ is the average work function of sample and tip, and $\hbar$ is Planck's constant divided by $2\pi$. Changing $s$ by 1 Å will thus produce an order-of-magnitude change in the current. A fixed value of tip-sample separation is achieved by opening the feedback loop of the STM (which normally, when closed, maintains a tip height such that the tunnel current in constant), and then scanning the voltage permits acquisition of a spectrum. This procedure can be repeated at different values of tip-sample separation, by adding a small offset to the tip height prior to acquisition of the spectrum (or by using a different sample-tip voltage for the constant-current condition prior to opening the feedback loop, which has the effect of establishing a different tip-sample separation).

For relatively small values of tip-sample separation, spectra can be acquired that accurately probe band-edge features, but the voltage range of such measurements must be limited in order to prevent the current from being too large. Subsequent measurements at greater tip-sample voltages can be made, with an increased voltage range for each such measurement (still taking care to prevent currents that are too large, maintaining the current to be ≲1 nA). In order to display such measurements all together on a single plot with relative magnitudes that are meaningful [7], it is convenient to normalize the spectra by multiplying them by an appropriate exponential term. We assume that $\kappa$ is independent of voltage and given by a value $\bar{\kappa}$. Then, with a fixed $s$ value for each spectrum, we have

$$I(V,s) = g(V)e^{-2\bar{\kappa}s} \tag{1a}$$

$$\frac{dI(V,s)}{dV} = \frac{dg(V)}{dV}e^{-2\bar{\kappa}s} \quad . \tag{1b}$$

Consider a spectra measured with some arbitrary value of $s$, which we want to normalize to a specified separation, $s_1$. This can be accomplished by multiplying the data acquired at separation $s$ by a factor $\exp(2\bar{\kappa}\Delta s)$,

$$I(V,s_1) = I(V,s)\,e^{2\bar{\kappa}\Delta s} \tag{2a}$$

$$\frac{dI(V,s_1)}{dV} = \frac{dI(V,s)}{dV}e^{2\bar{\kappa}\Delta s} \quad . \tag{2b}$$

where $\Delta s = s - s_1$ is the change in separation relative to some nominal reference point $s_1$, e.g., the separation at constant-current prior to when the measurement sequence began. For example, if the tip is moved *closer* to the surface, we have $\Delta s < 0$; such spectra will have relatively large magnitude, and hence the normalization *reduces* their magnitude such that it will be in approximate correspondence with other spectra of the same set (acquired with various $\Delta s$ values) that are normalized according to Eq. (2). It is in the essential nature of STS measurements that such a data set be measured in a relatively short amount of time, typically less than a minute, such that thermal drift (and piezoelectric creep) effects do not cause the tip do vary its relative position (laterally or vertically) over the surface. Acquisition of spectra using an instrument that operates at low temperature minimizes drift effects, but nonetheless, rapid data acquisition is still desirable.

Several issues arise in the *normalization to constant-z* represented by Eq. (2). The first is determining a suitable value of $\kappa$ to use. There are many effects that can produce actual $\kappa$ values that differ from the "ideal" value of 1.1 Å$^{-1}$ mentioned above (some of these effects are discussed later in this paper), so in principle, $\kappa$ can have some dependence both on voltage and on tip-sample separation. In using Eq. (2), we are assuming a single $\kappa$ value, $\bar{\kappa}$, often determined by a few current vs. separation measurements at different voltages. One can thus display a set of current vs. voltage ($I$ vs. $V$) measurements, or better still conductance vs. voltage ($dI/dV$ vs. $V$) measurements, acquired with various (known) $\Delta s$, all together on a single plot. Often a logarithmic scale will be used for the current or conductance, to allow presentation of the data over several orders of magnitude.

The voltage $V$ refers to the sample bias voltage relative to the probe-tip, with normally empty (CB) states generally probed with large positive voltages and normally filled (VB) states generally probed with negative voltages. Neglecting for a moment any effects of tip-induced band bending (TIBB) [7], i.e. when some of the electric field across the vacuum junction extends into the semiconductor itself, then the relationship between the sample bias voltage $V$ and the energy of a state in the sample relative to the sample Fermi energy, $E - E_F$, is given by

$$eV = E - E_F \tag{3}$$

where $e$ is the elementary charge (a positive quantity). Thus, we have the very convenient relationship that $V$ expressed in volts is numerically equal to $E - E_F$ expressed in electron-volts, and hence the measurements we are performing are called "spectroscopy" (and the measured data are called "spectra") since we are scanning over the energy states of the sample.

For semiconductors, it may happen that TIBB effects are not small, thus producing errors in the application of Eq. (3). Such effects can be estimated, and the voltage scale of the data appropriately corrected [7,22,23,24,25]. A bigger effect occurs when states of the semiconductor that are normally empty are pulled down (due to TIBB) to be below the sample Fermi energy, thus becoming filled with electrons and contributing to the tunnel current (or conversely, normally filled states being pulled up and becoming empty). We can then say that an "accumulation layer" exists on the surface, involving occupation of the CB or VB (and/or changes in occupation of discrete levels, i.e. arising from defects, on the surface) [7, 26]. A number of nonobvious effects then occur, associated with the changes in occupation of states due to their energy position changing with respect to the Fermi energy of the sample (rather than of the tip, as is implicitly assumed in Eq. (3)) [26,27].

A further complication in this situation is that, in early work, the phrase "accumulation layer" (or "accumulation current") was avoided, since, for the first system in which this was observed (cleaved GaAs(110) surface), there was a strong restriction in the magnitude of the accumulation current [7], for the CB in particular. This restriction was so strong that, indeed, significant accumulation did *not* occur; rather, only a weak (but clear) signal of such current was observed, with magnitude that was approximately equal to what was expected from CB doping of the material *in the absence* of TIBB. Hence, this observed current was given the somewhat ungainly name of "dopant-induced current" [7] (the TIBB in the semiconductor was not assumed to be absent, but rather, it was thought that, perhaps, refilling of accumulation layer states was sufficiently slow so that a sustained current through such states was not possible). Resolution of this puzzle did not occur until 22 years later [28], when it was finally recognized that surface states on the GaAs(110) surface act to greatly restrict the TIBB. By that time, clear accumulation contribution to the tunnel current had been observed in other systems [26,29]. Hence, all of the early usages of the term "dopant induced" can be equivalently called "accumulation", while recognizing for the GaAs(110) CB that this accumulation is very restricted, due to the surface states (located in energy just above the CB edge).

There are several other issues that arise when applying the normalization to constant-z, as in Eq. (2), to data sets consisting of several spectra acquired with different values of $\Delta s$ (i.e. a separate, fixed value of $\Delta s$ used for each spectrum). If we normalize all of the spectra to constant-z using Eq. (2) and plot them one atop the other, they may not exactly line up with each other in the overlap intervals of the respective spectra. A fundamental reason for this lack of alignment is the (weak) voltage- and/or separation-dependence of $\kappa$. An additional reason is that thermal drift or piezo creep cause the probe-tip to actually move slightly during the measurement sequence (additionally, the tip itself may change its characteristics, an inherent issue during STS measurements and one that can only be detected by a large number of repeated measurements, i.e. to check for reproducibility).

Another issue associated with the use of Eq. (2), one that is relatively minor but nonetheless bothersome, is that if we want to plot the data on a logarithmic scale in order to properly view all orders of magnitude in it, we must strip off the portions of the data that are limited by instrumental noise. That is to say, each spectrum will have some region, typically equal to or larger than the band gap (but varying in size depending on the proximity of the tip to the surface, and also affected by the above-mentioned accumulation current effects) where the current or conductance is limited by noise. On the logarithmic plot, this noise can appear at all values with magnitude less than the noise level, i.e. in principle extending over a large portion of the plot. Such noise prevents clear viewing of one spectrum on top of another (since the noise can completely obscure a spectrum measured with reduced tip-sample separation, but then normalized such that it appears at relatively small current values). It is not a problem to strip off the noise, since a measurement of the instrumental noise level can easily be made, and the noise-limited voltage interval for any given spectrum can be determined in a straightforward manner [7]. However, this process demands some attention to each and every spectrum individually, something that is rather time consuming. It is desirable to achieve a more automated means of acquiring and displaying the data.

One method to acquire and display the data in a more automated way, while still achieving high dynamic range, is to *continuously* vary the tip-sample separation during the measurement. One means of doing so is to linearly vary the tip-sample separation in accordance with the magnitude of the voltage [8,30]. It is convenient to permit different slopes for this variation on the positive-and negative-voltage sides of the voltage scan, with these values chosen for convenience during each experiment such that the magnitude of the current and conductance is limited to be below some value ($\lesssim 1$ nA). We acknowledge that some workers may consider a *continuously* varying tip-sample separation to be deleterious to the measurement, i.e., in the sense that it significantly distorts the data. However, it must be realized that since $\kappa$ itself has some dependence on voltage, even for a measurement at *constant* tip-sample separation, the tunnel barrier will be changing significantly (a small change in $\kappa$ producing a relatively large change in the transmission through the barrier).

By varying the tip-sample separation with a V-shaped ramp, i.e. moving the tip towards the surface as the magnitude of the voltage is reduced to zero, and then moving it away from the surface as the magnitude of the voltage is increased (for voltages with opposite sign as the initial voltage of the scan), we can achieve a great increase in the current and conductance. We thus can amplify their values to be well above the noise level, while "distorting" the data only in the relatively well-known way as given by Eq. (1). It should also be noted that, in most measurements, the magnitude of the current or conductance is not the primary thing that is studied in a spectrum. Rather, it is the voltage (energy) of given spectral features, so that some modest "distortion" on the magnitudes is not a problem. For any experiments in which variation in magnitude *are* important, then they of course must be performed with identical V-shaped ramps for the variation in tip-sample separation (as well as identical tip conditions and overall tunneling parameters; hence, inherently, comparative measurements of this type are performed one right after the other, without change in any parameter aside from e.g. the spatial location at which the two spectra are being acquired).

To write an expression for the applied variation in tip-sample separation, we first note that the measurement might proceed from positive to negative voltages, or from negative to positive voltages (whichever is most convenient during the experiment, i.e., since sometimes performing constant-current scans on a given surface with a given tip is most stable using a particular sign of the sample bias voltage). Let us say that the scan proceeds from $V_1$ to $V_2$, where $V_1$ and $V_2$ have opposite signs but we are not specifying which is positive and which negative. In the first part of the scan, the voltage varies between $V_1$ and 0, and in the second part between 0 and $V_2$ (actually, data is often acquired, as well, in the return direction from $V_2$ to $V_1$, with an obvious replication of parameters below for this reverse scan). As a reference point for the separation, we can use $s_1$, the separation for the constant-current imaging that takes place prior to the spectroscopy measurement. Additionally, we also often include an offset to that value, $\Delta s_1$, applied so as to either produce an overall increase or decrease in the magnitude of the current and conductance. The $\Delta s(V)$ variation for the V-shaped ramp can thus be written

$$\Delta s(V) = \begin{cases} \Delta s_1 - a_1|V_1 - V| & \text{first part of scan} \quad\quad (4a) \\ \Delta s_1 - a_1|V_1| + a_2|V_2 - V| & \text{second part of scan} \quad\quad (4b) \end{cases}$$

where, as described above, $\Delta s_1$, $a_1$, and $a_2$ are chosen for reasons of convenience in the measurements. The parameters $a_1$ and $a_2$ typically have chosen values of $1-2$ Å/V, large enough to produce significant magnification of the current and conductance at low voltages near 0 V, but not too large so as to produce tip-sample separations that are too small and hence lead to nonideal tunneling conditions. With both $a_1$ and $a_2$ being *positive*, then in Eq. (4a), we have $|V_1 - V| > 0$ and $s(V)$ is *decreasing* during the first part of the scan, so a minus sign necessarily precedes the $a_1|V_1 - V|$ term in Eq. (4a). Alternatively, in Eq. (4b), we also have $|V_2 - V| > 0$ but now $s(V)$ is *increasing*, so a plus sign necessarily precedes $a_2|V_2 - V|$.

Concerning normalization of data acquired with a V-shaped ramp for $\Delta s(V)$, we can still utilize a formula analogous to Eq. (2), but we must reinterpret the conductance there, $dI/dV$, to be the *partial* derivative of the current measured at constant tip-sample separation, $(\partial I/\partial V)|_s$. This is the quantity that is measured by a lock-in amplifier during the experiment (using a modulation frequency of typically 1 kHz). Even if the tip-sample separation is slowly varied during the scan, then this conductance at constant separation is still measured, albeit it at a different tip-sample separation for each value of voltage. With a continuously varying $s(V)$, we now must replace Eqs. (1) and (2) by

$$I(V, s(V)) = g(V)e^{-2\bar{\kappa}s(V)} \tag{5a}$$

$$\left.\frac{\partial I(V, s(V))}{\partial V}\right|_s = \frac{dg(V)}{dV}e^{-2\bar{\kappa}s(V)} \tag{5b}$$

and

$$I(V, s_1) = I(V, s(V))\, e^{2\bar{\kappa}\Delta s(V)} \tag{6a}$$

$$\left.\frac{\partial I(V, s_1)}{\partial V}\right|_s = \left.\frac{\partial I(V, s(V))}{\partial V}\right|_s e^{2\bar{\kappa}\Delta s(V)} \tag{6b}$$

with $\Delta s(V)$ given by Eq. (4).

An example of such normalized STS data is displayed in Fig. 1, for a monolayer of WSe$_2$ [9,10] as well as from an exposed area of EG that the WSe$_2$ film lies on. Results are shown for several different values of the ramps for $\Delta s(V)$; e.g., from the values listed in the caption, we have for curve (a): $\Delta s_1 = -0.5$ Å, $V_1 = -2.1$ V, $V_2 = 2.2$ V, and ramp values of $a_1 = 2.5/2.1 \cong 1.19$ Å/V and $a_2 = 2.8/2.2 \cong 1.27$ Å/V. Note the increase in dynamic range of the data as the ramp values are increased, i.e., with curve (a) clearly revealing background conductance observed within the WSe$_2$ band gap (between about $-0.8$ and 1.2 V). This conductance within the band gap is interpreted as arising from the graphene located underneath the WSe$_2$ layer, as further discussed in Section III. The noise level of each measurement in Fig. 1 is marked on the spectra, with this quantity determined simply by performing a noise measurement with the probe-tip pulled back from the surface (such that the current is zero, aside from the noise), and then normalizing that measured noise value in precisely the same manner as for the respective spectra. Of course, the noise level decreases as the ramp values increase. The various observed features in VB and CB are marked in accordance with their presumed origin, based on band structure computations and angle-resolved photoemission spectroscopy (ARPES) [31,32,33], as discussed in more detail in Section III. Note that all of the spectra in Fig. 1 clearly reveal the K$_V$ band in the VB; this band was also seen in the STS spectra of Yankowitz et al. [16], but not directly revealed in spectra of Hill et al. [21] and Zhang et al. [34] due to limited dynamic range (although the latter work also performed measurements with varying $s(V)$, employing a constant-current condition, as further described below). An increased dynamic range is evident in our data, due to a reduced noise level in the data (i.e. better electrical shielding in the STM wiring and/or a better pre-amp) together with our use of the applied variation in $\Delta s(V)$.

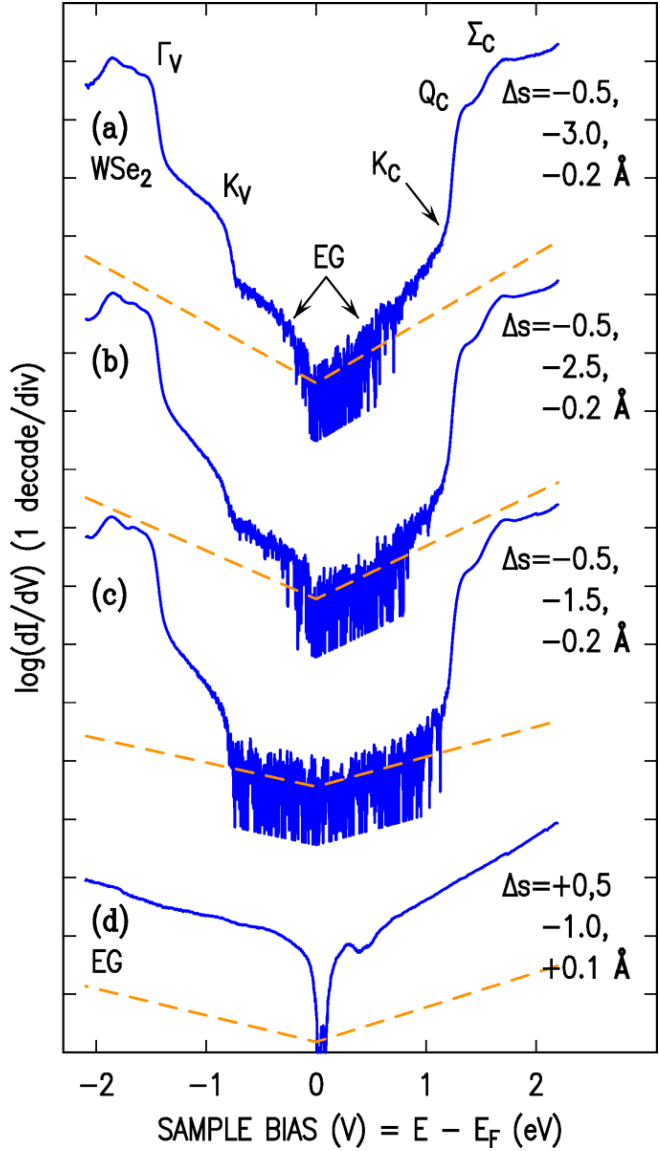

FIG 1. STS acquired at 5 K showing spectra from (a) – (c) a single monolayer of $WSe_2$ in 3 different measurements, and (d) epitaxial graphene (EG). The applied variations in $\Delta s$ at the starting voltage of the scan ($-2.1$ V), at 0 V, and at the ending voltage ($+2.2$ V), respectively, are listed. A linear variation in $\Delta s(V)$ is utilized between 0 V and each of the starting or ending voltages. The noise level for each measurement is shown by the thin dashed lines. Various features in the $WSe_2$ spectra are labelled in accordance with their presumed origin from bands centered around wave-vector values at the $\Gamma$, K, Q and $\Sigma$ points within the Brillouin zone, with subscript "V" denoting valence band and "C" conduction band. A background signal in the $WSe_2$ spectra is indicated, from EG that lies below the $WSe_2$ monolayer.

One important aspect for performing measurements with varying tip-sample separation, is that the conductance really should be measured during the scan using a lock-in amplifier, rather than taking a numerical derivative of the data after the acquisition. It's not unusual to use a lock-in method, of course, although depending on the precise details of the experiment and the data acquisition, it may be that simply measuring current vs. voltage is preferred (e.g. if very rapid data acquisition is required). In this case, when only current vs. voltage is measured, variation of the tip-sample separation during the measurement should be avoided, since it would then be impossible to obtain a realistic measure of conductance.

To make this situation clear, we return to the discussion preceding Eq. (5), where it was noted that the conductance, as usually defined, is the *partial* derivative with respect to voltage at *constant* tip-sample separation, $(\partial I/\partial V)|_s$. This quantity differs from what we would obtain if we take a *total* derivative to the measured current vs. voltage, $dI/dV$, according to

$$\frac{dI}{dV} = \frac{\partial I}{\partial V}\bigg|_s + \frac{\partial I}{\partial s}\bigg|_V \frac{ds}{dV} \quad . \tag{7}$$

Of course, it is the total derivative $dI/dV$ that is obtained if a numerical derivative of the current is performed. If the tip-sample separation is not being varied, then there is no difference between $dI/dV$ and $(\partial I/\partial V)|_s$, i.e. the second term on the right-hand side of Eq. (7) is zero. But, when the tip-sample separation is being varied, e.g., using a V-shaped ramp so that $ds/dV$ has a constant value for a given sign of the voltage, then the second term on the right-hand side of Eq. (7) can be substantial in magnitude compared to the first term. For this reason, the numerical derivative of the current, $dI/dV$, will differ from the conductance, $(\partial I/\partial V)|_s$, when the tip-sample separation is varied.

The relationship expressed by Eq. (7) can actually be used to advantage, to gain an approximate measure of $(\partial I/\partial s)|_V$, and hence $\kappa$, from any given spectroscopy measurement performed with varying tip-sample separation. The detailed method has been described previously [30]. Such a measure of $\kappa$ is "approximate" in the sense that it's a rather crude measurement compared to what we might obtain directly from a measurement of current vs. tip-sample separation (at given sample bias voltage). Numerically, including effects of noise in the data, it is actually relatively difficult to use Eq. (6) to obtain $\kappa$ in this manner (since a numerical derivative is necessary for obtaining $dI/dV$, and furthermore, a difference between that term and the partial derivative $(\partial I/\partial V)|_s$ must be taken), but nonetheless an approximate measure of $\kappa$ can indeed be obtained.

A much better means of determining $\kappa$, and in particular its voltage dependence, is to directly measure $\partial I/\partial s$ at constant $V$. This is often done at constant current, thus maintaining the current to be relatively large (i.e. whatever value is chosen for the constant-current set-point). A recent work by Zhang et al. presents measurements of this type, performed on a monolayer of $WSe_2$ [34]. Important conclusions concerning the identification of the bands from which specific contribution to the current and conductance originated was enabled by these quantitative measurements of $\kappa$. In particular, to lowest order, the variation in $\kappa$ with the parallel wave-vector, $k_{||}$, of an electronic state is given by [35,36,37,38]

$$\kappa = \sqrt{2m_0\bar{\phi}/\hbar^2 + k_{||}^2} \quad . \tag{8}$$

Thus, measurement of $\kappa$ can be used to infer a value of $k_{||}$, from which the states contributing to the current and conductance can be identified as arising from some specific band (valley), e.g. centered about some nonzero $k_{||}$ value. However, a danger in such measurements is that the probe-tip might come *too* close to the surface, leading to non-ideal tunneling conditions (collapse of the tunnel barrier) and hence making it difficult to directly interpret the observed $\kappa$ values as pertaining to a specific electronic band of the sample. We return in Section IV to further discuss this situation, in connection with the results of Zhang et al. [34].

It is important to note that two other, powerful methods exist to determine $k_{||}$. The first, based on STM/STS, is to use utilize Friedel oscillations as observed in scattering from defects, with the deduced wavelength of the oscillation providing a measure of $k_{||}$ [16,39]. Second, much more generally, ARPES permits determination of the band structure associated with filled electronic states, allowing determination of $k_{\parallel}$ in a manner that is much more direct and reliable than any STM/STS method [32,33]. Of course, ARPES lacks the high spatial resolution associated with STM/STS, but nonetheless, recent advances in ARPES with μm-scale resolution are most notable [32,33].

A completely separate means of normalizing tunneling spectra, different than Eq. (2) or (6), is to form the so-called normalized conductance,

$$\frac{dI/dV}{I/V} \qquad \text{or} \qquad \frac{dI/dV}{\overline{I/V}} \quad . \tag{9}$$

The difference between these two forms, in principle, is that some "broadening" has been applied to the quantity $I/V$ in the denominator of the right-hand expression (thus forming $\overline{I/V}$), whereas no such broadening has been applied to $I/V$ in the denominator of the left-hand expression. Actually, in practice, it is very difficult to compute $I/V$ without including some broadening (intentional or unintentional), as previously discussed [40]. In any event, for the right-hand form in Eq. (9), the broadening is performed over a relatively large voltage range, sufficient to span the band gap, such that $\overline{I/V}$ forms a suitable quantity with which to normalize $dI/dV$ [40]. Another aspect of the expressions in Eq. (9) that is important to note is that the quantity $dI/dV$ written there actually signifies the *partial* derivative $(\partial I/\partial V)|_s$, measured at constant $s$. (In early application of this normalized conductance, measurements were indeed performed at fixed tip-sample separation, so there was no difference between $dI/dV$ and $(\partial I/\partial V)|_s$. In later applications, when the separation was varied continuously with the sample voltage, the terminology for referring to normalized conductance was left unchanged from Eq. (8), even though it would have been more precise to use $(\partial I/\partial V)|_s/(I/V)$ or $(\partial I/\partial V)|_s/(\overline{I/V})$ for labelling the normalized conductance, rather than continuing to use the terminology of Eq. (8)).

The *normalized conductance* form of Eq. (9) has been found to be quite useful in displaying STS data, particularly data acquired from semiconductors with band gaps on the order of ~1 eV in size [8,17,20,30,40,41]. In this form, clear delineation of the band edges as well as features well inside the bands (and in the band gap, if such features occur) is accomplished. In some cases, display of data using *both* of the normalized forms given by Eqs. (2) and (9) is useful [42]. Care is required in interpretation of STS data, since the manner in which it is plotted can end up emphasizing certain features at the expense of others (e.g. compare Refs. [17] and [19]).

In this regard, a common misconception in the literature is that the normalized conductance of Eq. (9) somehow corresponds directly to a local density of states. This interpretation is an overstatement as to the usefulness of the forms in Eq. (9). Rather, it is best simply to view this type of normalization of the spectra as a qualitative means of permitting a convenient view of the data – indeed, one that can be accomplished on a linear scale and still reveal all significant features of the data. On the other hand, it must be acknowledged that this type of normalization is somewhat qualitative, in the sense that the details of the procedure for computing $(dI/dV)/(\overline{I/V})$ can affect the outcome. This normalization should best be viewed as a type of "background subtraction", but performed in a multiplicative manner using a background constructed in a somewhat complicated way. In this way, all features in the data are clearly apparent, but the relative magnitudes of one part of the spectrum compared to another are somewhat dependent on the normalization procedure. As such, we consider the normalized conductance quantities of Eq. (9) to be unsuitable for simulation. That is to say, one might employ a band edge or peak location in such normalized spectra in a relatively reliable manner, but fitting of an entire spectrum to a theoretical form is not possible unless that theoretical form *also* includes all details of the normalization procedure.

Hence, for detailed curve fitting of spectra, it is far preferable to simply utilize the normalization to constant-z, as in Eq. (2); this was the procedure followed, e.g., in the extensive analysis of Ref. [24]. In such simulations, the V-shaped ramp utilized in the data acquisition is included, i.e., the simulation is performed along precisely the same $\Delta s(V)$ curve used for the measurement. For viewing the comparison between experiment and theory, both the experimental data and the simulated curves are normalized to constant-z, utilizing Eq. (2) and using the same value of $\kappa$ (typically about 1 Å$^{-1}$) for both theory and experiment.

## B. Simulation

In this Section, we discuss a second major aspect of STS analysis (the first being the "normalization" discussed in the prior Section), namely, simulation of spectra in order to enable curve fitting of a theoretical curve to a measured one. The goal of such curve fitting, for semiconducting materials, is to obtain a quantitative measure of parameters such as band-edge energies. As an initial step in this procedure, it is necessary to tentatively associate various features in the spectra with specific bands (i.e. valleys) of the VB or CB. For this purpose, consultation of existing band structures can be done, e.g. Refs. [31] or [32] (without, or with, spin-orbit interaction respectively). Then, detailed simulations are performed in order to verify these initial associations.

As discussed at length by Duke [43], there are two main methods for computing tunnel currents. The first is the "phenomenological approach", utilized in early work by Simmons [44], in which the current is computed by integrating over all states and including a term involving the velocity (momentum) at which a carrier is incident on the tunnel barrier. This type of method, generalized to permit dealing with semiconductor materials having band of specific effective mass, has been extensively utilizing for computations of TIBB and associated tunnel currents [7,22,23,24,25]. An advantage of this type of method is that it can be straightforwardly applied to situations in which a depletion layer occurs at the semiconductor surface, i.e., as important in the TIBB problem.

The second main approach for computing tunnel currents is the method of Bardeen [45], sometimes called the "Bardeen approach" (also referred to as the transfer-Hamiltonian approach). This is a perturbative treatment, in which an "overlap" of states associated with each tunneling electrode (i.e. states that are unperturbed by the presence of the other electrode), with suitable consideration of their relative occupations, is seen to form the tunnel current. The superiority of this picture compared to the phenomenological one is clear if we consider tunneling between a surface state (i.e. confined to the surface, with no momentum in the z-direction) and a probe-tip. The surface state, being by definition confined to the surface, has no momentum in the z-direction; hence, within the "phenomenological" picture it would contribute zero to the tunnel current. However, surface states were observed early-on in STM/STS experiments to make a large contribution to the tunnel current [3], thus making it clear that such states do indeed contribute to the tunnel current.

Tersoff and Hamann applied the Bardeen approach to the scanning tunneling microscope, i.e., having a sharp probe-tip as one of the electrodes in the tunneling configuration [35,37,38]. Their theory was largely developed to explain the spatial resolution of the STM, and in this regard it was very successful (a competing theory at the time, based on scattering [36,37], may in fact be more accurate, but it lacks the easy and intuitive viewpoint offered by the Bardeen approach). In the present work, since we are concerned with a 2D material (not having any momentum in the z-direction), it is clear that we must employ the Bardeen approach rather than the phenomenological approach for computing tunneling currents. We therefore seek to apply the theory of Tersoff and Hamann (hereafter denoted as TH). However, their work was applied only to low bias voltages, so we must slightly generalize it such that it also can be applied to finite bias voltages, as large as a few V. For this purpose, following their work, we re-derive a number of results.

As a starting point, we consider the current when tunneling from the tip (subscript $t$) into states of a given band of the sample (subscript $S$) [46], according to Bardeen's approach,

$$I = \frac{4\pi e g}{\hbar} \sum_{\alpha,\beta} |M_{\alpha,\beta}|^2 \left[ f_t(E_\alpha) - f_S(E_\beta) \right] \delta(E_\alpha - E_\beta) \tag{10}$$

where $\alpha$ and $\beta$ label the states of the tip and sample, respectively, having energies $E_\alpha$ and $E_\beta$, $f_t(E_\alpha)$ and $f_S(E_\beta)$ are the respective Fermi occupation factors, $M_{\alpha,\beta}$ is the "matrix element" for the process, and $g$ is the degeneracy of the band being considered. Denoting the Fermi energy of the sample by $E_F$ and that of the tip by $E_F + eV$, the Fermi occupation factors are given by

$$f_t(E_\alpha) = \frac{1}{1 + e^{-(E_\alpha - E_F - eV)/kT}} \tag{11a}$$

$$f_S(E_\beta) = \frac{1}{1 + e^{-(E_\beta - E_F)/kT}} \tag{11b}$$

for temperature $T$ and where $k$ is Boltzmann's constant. The term $f_t(E_\alpha) - f_S(E_\beta)$ provides a "window" for tunneling, with significant current only occurring for states having $E_\alpha = E_\beta$ that lies between Fermi energies of tip and sample or within a few $kT$ of that range. The matrix element $M_{\alpha,\beta}$, in general, is given by

$$M_{\alpha,\beta} = \frac{\hbar^2}{2m_0} \int d\mathbf{S} \cdot \left( \psi_\alpha^* \nabla \psi_\beta - \psi_\beta \nabla \psi_\alpha^* \right) \tag{12}$$

where $\psi_\alpha$ and $\psi_\beta$ are wave functions of tip and sample, respectively, and where the integral is evaluated over any surface separating the two electrodes.

Following TH, we consider a sharp, metallic probe-tip, and we express this matrix element in terms of a normalization volume of the probe-tip $\Omega_t$, a radius of curvature of the probe-tip $R$, and a value for the sample wave function. However, we do not evaluate the latter at the center of radius-of-curvature of the probe-tip (which is useful for understanding the resolution of the STM [38], but is somewhat unphysical in terms of spectroscopy measurements). Rather, we evaluate it at the apex of the probe-tip, thus absorbing a factor of $e^{\kappa R}$ from the TH treatment into our expression for the tip wave function. Denoting this point by $\mathbf{z}_0$, we thus find

$$M_{\alpha,\beta} = \frac{\hbar^2}{2m_0} \frac{4\pi R g}{\Omega_t^{1/2}} \psi_\beta(\mathbf{z}_0). \tag{13}$$

Inserting this expression into Eq. (12), and then evaluating Eq. (10) where we rewrite the sum over $\beta$ as an integral over states of the sample labelled by $\mathbf{k}_{\parallel}$ (and the sum over $\alpha$ is expressed, following TH, simply as a density of states of the tip $D_t$), we find for the current of a single band of sample states,

$$I = \frac{8\pi e R^2 \hbar^3 g A_S D_t}{m_0^2} \int d^2 \mathbf{k}_{||} \left| \psi_{\mathbf{k}_{||}}(\mathbf{z}_0) \right|^2 \left[ f_t \left( E_{\mathbf{k}_{||}} \right) - f_S \left( E_{\mathbf{k}_{||}} \right) \right] \tag{14}$$

where $A_S$ is a normalization area for the states of the sample.

Let us now consider applying such theories for STS to obtain detailed descriptions of experimental data, i.e., simulating the data using theoretical curves that can be fit to the data in order to extract parameters of interest from the data. In this regard, we encounter a significant problem (one which is common to many types of experimental data, not just STS) – the description of Eq. (14) requires full knowledge of the atomic and electronic arrangement in the sample, including at its surface. In many cases, such knowledge is not available at the outset of a study, i.e. obtaining such knowledge is one of the main goals of the study. Of course, one can always consider a range of possibilities for the atomic and electronic arrangements, and make detailed computations of the resulting wave functions, and hence the result of Eq. (14). However, here again we encounter a problem, in that realistic results for such computations are quite difficult (and very time consuming) to obtain. Despite the prevalence of "first-principles" computations using density-functional theory nowadays, it is well known the resulting band structures can still deviate substantially from what occurs in experiment, due to the approximate treatment of electron-electron interactions. Hence, we may well have available an approximate band structure with which to compare measured STS data to, but we should not assume that the various band-edge energies of this computed band structure are completely accurate. Rather, it is the goal of the experiment to determine such energies.

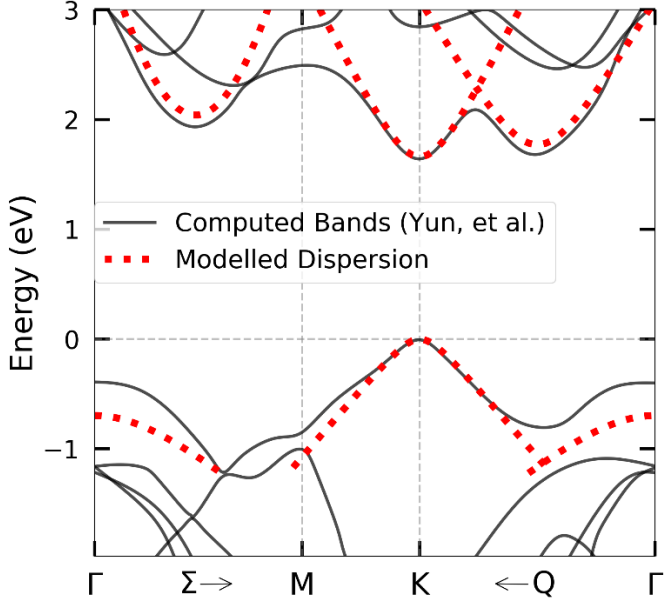

FIG 2. Computed band structure of WSe$_2$ by Yun et al. (Ref. [31]) as shown by thin black lines, compared to bands using hyperbolic dispersion as assumed in the present work for the purpose of fitting to the observed STS spectra. Results of this fitting can be seen in the locations of the onsets of the hyperbolic-shaped bands. For ease of viewing, an *additional* overall shift between CBs and VBs (revealed by the fitting) of 0.30 eV is not shown in the figure. Special points Γ, M, and K of the Brillouin zone are indicated, along with directions Q and Σ.

To illustrate these limitations in first-principles computations, we show in Fig. 2 a result of such computations by Yun et al. for WSe$_2$ [31], together with the locations of bands that we derive from our curve-fitting analysis of STS data as detailed in Section III. We should emphasize that our use of the results of Ref. [31] is only as an example; they are *not* state of the art in terms of computation results (since these early results did not utilize the hybrid density functionals, as discussed, e.g., in Ref. [32]), but they are nonetheless typical of what might be encountered when investigating a relatively unstudied material using STS. It is clear that a number of discrepancies occur between the predicted bands compared to the band onsets determined by our analysis of the measured STS spectra. These discrepancies all are due to inaccuracies of the first-principles theory used for those computations. Some such inaccuracies are well known, e.g. band-edge energies that deviate slightly from actual ones, due to neglect of parts of the electron-electron interaction term. (In Fig. 2 we have, for ease of viewing, shifted the overall separation between fit bands of the VB and CB, in order to approximately match the computed band structures. This shift, amounting to 0.30 eV in Fig. 2, is thus an *additional* discrepancy between experiment and theory that is not graphically represented in this plot). Additionally, there can be somewhat larger discrepancies between the fit bands and the computed ones, e.g. as shown for the Γ-band in the VB of WSe$_2$ in Fig. 2. In this case, the discrepancy arises from the relatively large spin-orbit interaction in WSe$_2$, as found in the energy splitting between the top of the Γ- and K-bands in the VB of WSe$_2$ revealed both by recent ARPES data as well as improved band structure computations [32].

Given that a band-structure computation that yields realistic energies for all bands is very challenging (i.e. not possible in a routine way), it is therefore desirable to use approximate models for the band structure of the sample in order to fit STS data. To this end, for the 2D materials that we are considering, we ignore all details of the sample wave functions other than their decay into the vacuum and their parallel wave-vector $\mathbf{k}_{||}$. Modeling the decay as it would occur in a trapezoidal barrier, we have

$$\left|\psi_{\mathbf{k}_{||}}(\mathbf{z}_0)\right|^2 \approx \frac{1}{A_S L} \exp\left\{-2s\sqrt{\frac{2m_0}{\hbar^2}\left(\bar{\phi} - \left(E_{\mathbf{k}_{||}} - E_F\right) + \frac{eV}{2}\right) + k_{||}^2}\right\} \tag{15}$$

where $L$ is a normalization length of the states of the 2D layer (approximately equal to the interlayer separation in a bulk crystal of the material).

We furthermore assume various models of bands (valleys) in the VB or CB. To introduce one such model, consider a band centered about the point, $\mathbf{k}_{||}^{(0)} = \left(k_x^{(0)}, k_y^{(0)}\right)$. In that case, one can write simply

$$E_{\mathbf{k}_{||}} = E_0 \pm \frac{\hbar^2|\mathbf{k}_{||} - \mathbf{k}_{||}^{(0)}|^2}{2m^*} \tag{16}$$

with $E_0$ being the onset of the band, $m^*$ its effective mass, and where the upper sign used for a CB ($E_{\mathbf{k}_{||}} \geq E_0$) and the lower sign for a VB ($E_{\mathbf{k}_{||}} \leq E_0$). Considering such a band, and including consideration of the periodic potential in the crystal, then a wave function associated with a state labelled by $\mathbf{k}_{||}$ will have Fourier components at all values of $\mathbf{k}_{||} + \mathbf{G}$, i.e. for all reciprocal lattice vectors $\mathbf{G}$. In other words, as in a "periodic zone scheme" [47], bands exist centered about all $\mathbf{k}_{||}^{(0)} + \mathbf{G}$ values. The magnitudes of the associated Fourier components can be substantial for the first few $\mathbf{G}$ values [38]. However, for terms with $\mathbf{G} \neq (0,0)$, the corresponding $k_{||}^2$ terms in the expression for $\kappa$ in Eqs. (8) and in the exponent of Eq. (15) should be replaced by $|\mathbf{k}_{||} + \mathbf{G}|^2$ [36,38]. Terms with the smallest values of $|\mathbf{k}_{||} + \mathbf{G}|^2$ will dominate in the current, since they decay relatively slowly from the surface. Hence, the integral over $\mathbf{k}_{||}$ in Eq. (14) can be restricted to the first Brillouin zone, and bands are included for all $\mathbf{k}_{||}^{(0)} + \mathbf{G}$ values that lie within, or on the edge, of this zone.

As further discussed in Section III, we find that the parabolic dispersions produced by Eq. (16) do not realistically match the known band structures of the materials we study. A somewhat better match is obtained using hyperbolic dispersion curves. We define these according to

$$E_{\mathbf{k}_{||}} = E_0 \pm v^2 m^* \left[ \sqrt{1 + \frac{\hbar^2 |\mathbf{k}_{||} - \mathbf{k}_{||}^{(0)}|^2}{(vm^*)^2}} - 1 \right].$$ (17)

Near the onset of the band, parabolic behavior is still obtained (as in Eq. 16)), but far from the onset the dispersion becomes linear. The slope of this linear part of the dispersion is given by $\hbar v$, i.e. this slope is determined by the parameter $v$, which has units of velocity.

With the hyperbolic bands of Eq. (17), together with the expression for the tunnel current in Eq. (14), we produce simulated conductance curves that we can compare to measured spectra. An additional effect that must be added to these simulated curves is the instrumental broadening (resolution) arising from the modulation voltage and low-pass filter of the lock-in amplifier used for the measurement. For the data of Fig. 1, a peak-to-peak modulation voltage of $V_{pp} = 50$ mV was employed, corresponding to a root-mean-squared (rms) value of $50/(2\sqrt{2}) = 17.7$ mV. This value corresponds to the standard deviation in a distribution that "broadens" the simulated data, with the variance of the distribution given by $\sigma_m^2 = (17.7 \text{ mV})^2$. (The use of the term "broaden" here is much different than at the end of Section II(A); there, we were discussing an assumed, phenomenological broadening of $I/V$ in order to produce a normalization term for $dI/dV$, whereas in the present case we are referring to an actual, physical broadening of the $dI/dV$ spectra). With knowledge of the $RC$ time constant of the low-pass filter, along with the rate of the voltage scan used for data acquisition, a variance for the low-pass filter if found to be $\sigma_f^2 = (2.8 \text{ mV})^2$. The instrumental broadening can thus be computed by employing Gaussian broadening, i.e. convoluting a simulated spectrum with the distribution $\exp[-(V'-V)^2/(2\sigma_b^2)]/(\sqrt{2\pi}\sigma_b)$, where $\sigma_b^2 = \sigma_m^2 + \sigma_f^2$. As we will see in Section III, a substantial amount of *additional* broadening is needed in order to explain the STS data; we incorporate this additional amount simply by adding another term to the sum that forms the total broadening

$$\sigma_b^2 = \sigma_m^2 + \sigma_f^2 + \sigma_a^2$$ (18)

where $\sigma_a$ refers to the additional broadening, i.e., beyond the instrumental effects. (The broadening of the computed spectra can either be performed in two steps, first using variances of $\sigma_m^2 + \sigma_f^2$ and then $\sigma_a^2$, or in a single step by using a variance of $\sigma_m^2 + \sigma_f^2 + \sigma_a^2$, and precisely the same result is found by either method).

We now return to consider, in a more general sense, models that have been employed for simulation of tunneling. In addition to the phenomenological and the Bardeen approaches, a third method is commonly utilized (e.g. by Hill et al. [21]), but it must be recognized as being rather approximate. In this method, the current is assumed to be proportional to an integral over energy of the form

$$I \propto \int_{-\infty}^{\infty} dE \, [f_t(E) - f_S(E)] \, \rho_S(E) \, T(E,V)$$ (19)

where $\rho_S(E)$ is an energy-dependent density of states of the sample (with any energy dependence of the tip density of states being neglected), and $T(E, V)$ is a transmission term for tunneling of the electrons through the vacuum. The transmission term for a trapezoidal barrier would be very similar to Eq. (15) (disregarding the prefactor of $1/A_S L$ there), except for the $k_{||}^2$ term in the square-root of Eq. (15). As discussed above, this term will vary for different states in a band, leading to significant changes in the tunneling transmission. However, a method based on Eq. (19) does not permit inclusion of this dependence, since the transmission term is assumed to *only* depend on the energy (and the bias voltage). Rather, to employ Eq. (19), the transmission would have to be expressed by utilizing a form similar to Eq. (15), but using $k_{||}^{(0)} \equiv \left| \mathbf{k}_{||}^{(0)} \right|$ rather than $k_{||} \equiv \left| \mathbf{k}_{||} \right|$ in that form. That is to say, if the actual transmission of the carriers depends on some wave-vector that cannot be written simply as an energy (i.e., as occurs for any energy band with nonzero $\mathbf{k}_{||}^{(0)}$), then Eq. (19) cannot be used to realistically generate simulated curves for the tunneling current.

### III. Matching of Theory to Experiment – WSe₂ Results and Discussion

Let us now turn to the procedure whereby the simulated conductance curves are matched to the observed STS spectra (Fig. 1), in order to determine parameters associated with the simulated curves. We focus here on qualitative aspects of this matching, with the results of quantitative curve fitting (including error analysis) described elsewhere [48]. The number of assumed bands and the central wave-vector of each is chosen to match band structure computations for WSe₂ (Fig. 2) [31], using an assumed hyperbolic dispersion for the bands, Eq. (17). Thus, for each band, in addition to its central wave-vector, there are parameters for the effective mass as well as the band velocity (slope of the dispersion curve far from the band onset).

Additionally, there is an "amplitude" parameter for each band that describes the intensity of the current and conductance from that band compared to another band. We do not attempt to establish an overall magnitude for the current and conductance (although the nominal tip-sample separation $s_1$ prior to application of a ramp is taken to be 7.5 Å in the simulations, which is typical of a theoretical tip-sample separation value when image potential effects are neglected [49] as is the case in this work). Hence, the amplitude parameter is taken to be unity for one of the bands; we choose that "reference" band to be the $\Gamma_V$ band. Importantly, we expect the amplitude parameters for all other bands to be on the order of 1. That is to say, the prefactor of Eq. (14) is not expected to vary greatly from band to band. All factors in that prefactor are constant, expect for the valley degeneracy which is either 1 or 2 depending on the band. (Spin degeneracy is included in Eq. (14), although spin-orbit interaction will lift that degeneracy and, for certain bands, will introduce another factor of 2 variation in the intensity, depending if the band has large splitting, or not). As commented prior to Eq. (15), we have neglected all details of the wave functions on the surface other than their parallel wave vector. Following the discussion of Tersoff and Hamann [38], we do not expect the details of the sample wave function to produce a substantial variation in the current, i.e., since the sharp probe-tip prevents any orthogonality between its states and those of the sample. Hence, overall, we expect the amplitude factors of the bands to be on the order of unity, i.e., approximately lying between 0.2 and 5.

Results for matching the VB of the WSe₂ spectrum (curve (a) from Fig. 1) are displayed in Fig. 3, with parameters listed in Table I. There are two bands involved here, the highest being a K_V band and the lower lying one being $\Gamma_V$. Two procedures are employed for the theory: In the first, only instrumental broadening is applied to the data. Comparing theory to experiment, we see for the $\Gamma_V$ band in particular that the steepness of its onset in that computation is much greater than bserved in the experiment. Hence, we add additional broadening with standard deviation 37.9 mV, in order to achieve a match between experiment and theory for the shape of the $\Gamma_V$-band onset. For the case of the $K_V$ band, its shape is dominated by the dispersion of the band, i.e., centered about the K point. We can obtain a good match between experiment and theory for this band, with or without additional assumed broadening (i.e. using slightly different band parameters in either case), but in our fits we choose to maintain the same additional broadening as for the $\Gamma_V$ band, as listed in Table I.

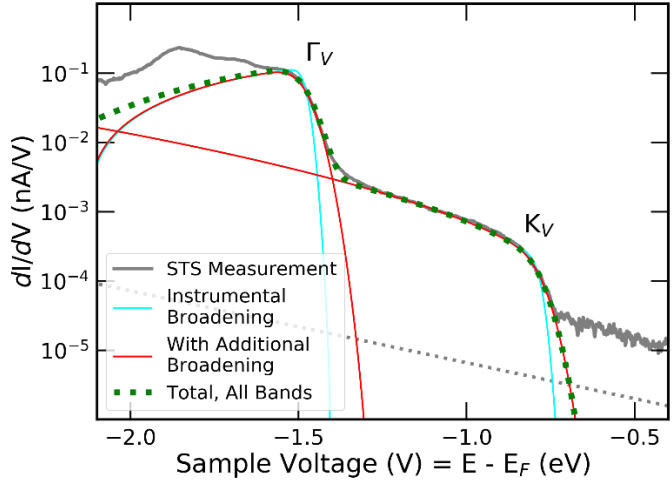

FIG 3. Matching of theory to experiment (gray line) for STS spectra of the VB of WSe₂. Three cases are shown for the theoretical curves, one with instrumental broadening only (cyan), one with additional broadening (red), and the total for all bands (green, dotted) of the curves from the individual bands that include the additional broadening.

TABLE I. Parameters for curve fitting of WSe₂ spectrum, listing for various bands the onset energy $E_0$ relative to the Fermi energy $E_F$, effective mass $m^*$ as ratio of the free-electron mass $m_0$, band velocity $v$, a scale factor for the amplitude, and additional broadening amount $\sigma_a$ (broadening for all VB, or all CB, are assumed to be equal). The locations of the $Q_C$ and $\Sigma_C$ bands are taken to be 0.52 and 0.47 of the $\Gamma$-K and $\Gamma$-M distances, respectively, following Ref. [31].

| band | $E_0 - E_F$ (eV) | $m^*/m_0$ | $v$ ($10^{15}$ Å/s) | scale | $\sigma_a$ (eV) |
|------|------------------|-----------|----------------------|-------|------------------|
| $\Gamma_V$ | −1.477 | 1.3 | 3 | 1 | 37.9 |
| $K_V$ | −0.78 | 0.2 | 4.1 | 3.0 | 37.9 |
| $K_C$ | 1.18 | 0.2 | 7 | 1.0 | 25.9 |
| $Q_C$ | 1.29 | 0.3 | 6 | 0.27 | 25.9 |
| $\Sigma_C$ | 1.56 | 0.3 | 20 | 0.25 | 25.9 |

The good match that we obtain between theory and experiment for the $K_V$ band, together with the fact that the amplitude factor listed in Table I for this band (relative to the $\Gamma_V$ band) is on the order of unity, gives us confidence that the theory we developed in Section II for describing these spectra does indeed work well. For comparison, in Supplementary Material we show results using parabolic bands, or neglecting the dispersion of the bands completely [50]; both of these computations produce relatively poor fits to the measured data. However, despite this success of our theory, the fact that we must employ 37.9 mV of additional Gaussian broadening (beyond the 17.9 mV of the instrumental broadening) in order to achieve a match to the $\Gamma_V$ band is something that is *not* contained within the theoretical description of Section II; clearly some additional effect in our physical description of the STS data is needed.

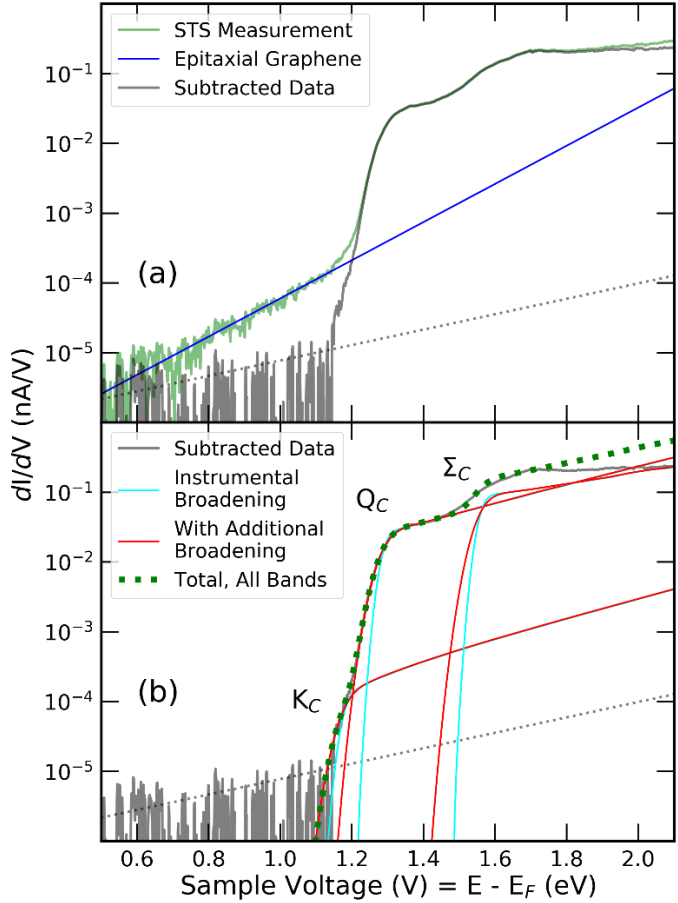

FIG 4. Matching of theory to experiment for STS spectra of the CB of WSe$_2$. (a) Background subtraction of conductance from EG (blue line) is shown, producing the background-subtracted data (grey line). (b) Curve fitting of background-subtracted data, using three cases for the theoretical curves, one with instrumental broadening only (cyan), one with additional broadening (red), and the total for all bands (green dotted) of the curves from the individual bands that include the additional broadening.

Let us now consider matching of theory to the CB of WSe$_2$, as shown in Fig. 4. There are three relevant bands as known from band structure computations: Two of these are relatively low-lying bands, one centered at the K point and the other at an intermediate point in the Brillouin zone along the Q direction (between $\Gamma$ and K). A higher lying band is centered around an intermediate point along the $\Sigma$ direction, between the $\Gamma$ and M. In the STS data for the CB, only two major onset features are observed, producing shoulders in the spectra at about 1.3 and 1.6 V. However, as just mentioned, near the lower 1.3-V onset there are actually expected to be two bands; we refer to these as K$_C$ and Q$_C$, with the upper band near 1.6 V denoted by $\Sigma_C$ (this latter band is the same as labelled by Zhang et al. as M$_C^*$ [34]). There is some controversy concerning the relative location of the K$_C$ and Q$_C$ bands, with a prior STS work placing Q$_C$ below K$_C$ [34], but a recent ARPES work finding, rather definitively, that K$_C$ is below Q$_C$ [33]. Importantly, the K$_C$ band, since it has substantially larger wave-vector than the Q$_C$ band, will correspondingly produce a significantly smaller conductance signal. Indeed, for the case of MoS$_2$, both the Q$_C$ and K$_C$ bands are clearly seen in STS data [9], with the conductance signal from the latter being about 2.5 orders-of-magnitude smaller than the former.

In the spectrum of Fig. 4(a), a background signal for the EG that lies below the WSe$_2$ monolayer limits our sensitivity to the possible presence of a signal from the K$_C$ band. Thus, we perform a subtraction of this background signal. As shown in Fig. 1(d), the EG itself has a conductance spectrum that is quite linear (on the semilog scale of the plot), particularly over the relevant range of 0.8 – 1.4 V. The background signal in the WSe$_2$ spectrum also appears to be quite linear over the first part of voltage range (up to nearly 1.2 V, above which the conductance from the Q$_C$ band turns on), although with a significantly different slope than that from the EG itself. This change in slope between a spectrum from the EG compared to a spectrum from EG *seen through* WSe$_2$ is expected. The WSe$_2$ acts as a tunnel barrier for the states of the EG, i.e., so long as those states do not overlap with energy states with the same value of parallel wave-vector from the WSe$_2$. This condition is indeed met for the energy range we are discussing. A moiré lattice forms between the graphene and the WSe$_2$, shifting the wave-vector of the relevant EG states from near the K point of EG to a location not far from the center of the WSe$_2$ Brillouin zone. There, the states exist in a band gap associated with the $\Gamma$ point of the WSe$_2$, the upper end of this gap being located at an energy well above the K$_C$ or Q$_C$ band of WSe$_2$. Hence, the states which contribute to the background signal from the EG can be viewed as simply tunneling through the WSe$_2$, with an energy-dependent decay length for this process (e.g. quite similar to that of Fig. 2(c) of Ref. [51], which shows states of graphene tunneling through h-BN). The energy dependence of this process then produces the change in slope for the EG background signal in the spectrum of Fig. 1(a), compared to the slope in the EG spectrum of Fig. 1(d).

Therefore, we consider it justifiable to simply fit the EG background of Fig. 4(a) with a line (on the semilog plot), extending beyond the onset of the Q$_C$ band, and then subtract this background contribution. The resulting background-subtracted data is shown in Fig. 4(b). Now, a small shoulder is apparent on the low-voltage side of the Q$_C$ band, and we associate this shoulder with the K$_C$ band. Fits are made to this data, using three bands as shown in Fig. 4(b) and with the resulting parameters listed in Table I. Most importantly, we find in our fitting procedure that the relative scale factors for the amplitudes of the K$_C$, Q$_C$, and $\Sigma_C$ bands are all of order unity. As described above, we view this requirement of having comparable contributions from all bands as being an essential ingredient in a proper description of the tunneling conductance. Hence, we believe that we have found an adequate theory with which to describe the STS data.

In contrast to the CB fitting results of Fig. 4(b) and Table I, we can instead assume, following Zhang et al. [34], that the Q$_C$ band lies below the K$_C$ band. It is straightforward to fit the data in this manner. However, the resulting scale factors for the amplitudes of the Q$_C$ and K$_C$ bands turn out to be 0.002 and 150, respectively, both of which deviate greatly from unity. On this basis, this type of fit can be rejected as being meaningful, and we conclude, in agreement with Nguyen et al. [33], that the K$_C$ band lies below the Q$_C$ band. As additional confirmation of our description of the relative magnitudes of tunneling from the Q$_C$ and K$_C$ bands, we note that Yankowitz et al. [16] clearly identified the former as being the one that dominates in the tunneling current, consistent with our interpretation.

Regarding the broadening needed to describe the CB spectra, as listed in Table I we find an optimal value of additional broadening for this data to be 25.9 mV, significantly less than for the VB data. We do not have an explanation for this difference between the results for the VB and CB, although we note that the additional broadening needed to fit the STS spectra also displays some variation with spatial location and/or probe tip. In Fig. S3 of the Supplementary Material, we show a spectrum acquired from a much different location on the surface (and with a different probe tip) than those of Fig. 1, and using in that case a smaller rms modulation voltage of 7.1 mV. Good fits are again obtained to the spectrum, with additional broadening of 33.9 and 43.9 mV for the VB and CB, respectively. Compared to the values in Table I, the former value is slightly smaller, but the latter value is substantially greater.

With the fits to both the VB and CB, a band gap for the monolayer $WSe_2$ can be determined: 1.96 eV, based on the separation of the $K_V$ and $K_C$ band onsets listed in Table I. An *estimated* uncertainty in this value is ±0.02 eV, based on inspection of the simulated curves compared to the measured spectra. This result for the band gap, 1.96 ±0.02 eV, is in agreement with that deduced in our prior work, where a separation between VB and CB of 1.93±0.02 eV was deduced from the raw data (i.e. simply by the apparent location of band edges, without any detailed fitting); correcting this value by adding the peak-to-peak value of the modulation voltage [9], 50 mV, yields 1.98±0.02 eV. (Note that the 50 mV peak-to-peak modulation is correctly reported in the present work, and similarly in Ref. [10], but it was *incorrectly* reported as a root-mean-square value in Ref. [9]).

Additionally, possible systematic effects in the measurements must be considered. One potential effect, as for any STS measurements involving semiconductors, is TIBB [7,22,23,24,25]. Since the $WSe_2$ studied here is lying on EG, then we might expect any such TIBB to be relatively small. Nonetheless it is important to determine, or at least estimate, just how small they are. Quantitative, numerical evaluation requires significant extension of existing computer codes [7,22,23,24,25], which we do not attempt here. Rather, we examine STS data acquired from monolayer $WSe_2$ lying on a different substrate. For the work of Yankowitz et al. [16], the $WSe_2$ was on a thick graphite flake. A band gap of 2.21±0.08 eV was deduced, directly from apparent band edge locations (without detailed fitting), which is substantially larger than our gap of 1.93±0.02 eV using the same, direct measurement, method. However, it is important to note that the EG used in our samples consists typically of 1 or 2 layers of graphene, below which is a carbon "buffer layer", and below that is the terminating layer of SiC which has dangling bonds extending up to the carbon buffer layer [39]. These dangling bonds form a partially filled band of states within the band gap of the SiC [39], and as such, they will act to significant constrain (reduce) any TIBB. We tentatively attribute the ~0.3 eV larger gap observed by Yankowitz et al., relative to our value, to TIBB in the $WSe_2$ plus underlying graphite. As for possible TIBB in our measurements, considering the properties of the EG and the underlying midgap dangling-bond states of the SiC, we feel that it would most likely be significantly less than this ~0.3 eV value, i.e., perhaps on the order of 20 – 50 meV.

Given this estimate of possible TIBB in our STS measurements, we note that this value is very similar in magnitude to the "additional broadening" that we find necessary to assume in order to produce good fits of the simulated spectra to the measurements. TIBB within the $WSe_2$ would tend to create a small, lateral "potential barrier" beneath the probe-tip, i.e., having height of approximately 20 – 50 meV and lateral extent on the few-nm range (determined by the probe-tip sharpness as well as screening from underlying EG and SiC). Some variation in such TIBB can be expected depending on probe-tip shape as well as spatial location on the sample (e.g. depending if there is 1 or 2 layers of EG on the SiC). More quantitative evaluation of the TIBB is clearly needed in order to determine whether or not this is indeed the source of the additional broadening that we find necessary to include in our simulations of the spectra. Nonetheless, we consider it quite possible that TIBB may account for the broadening of the spectra (and if so, then the "additional broadening" for the fits would have a detailed form that is different than a Gaussian, with the actual band onsets lying on one side of the broadened form, hence producing band gaps smaller than quoted above).

## IV. Conclusions

In this work, we have described the acquisition and analysis of STS data, with emphasis on spectra acquired with high dynamic range. We have developed a theory, based on the Bardeen approach as utilized by Tersoff and Hamann [35,38], for describing spectra acquired from 2D materials. A proper treatment of the parallel wave-vector is an essential aspect of our theory. Bands are modeled as having hyperbolic dispersion, with effective mass and band velocity (slope of dispersion curve far from a band extremum) chosen to match the band structure of the 2D material as known from computation. We have applied this theory to fit observed spectra acquired from a monolayer of $WSe_2$, residing on EG. Good fits between experiment and theory are obtained, and identification of various bands (consistent with prior band-structure computations [31,32]) is made.

Our results reveal band-onset energies of the VB that, in comparison with first-principles computations [31,32], are seen to be are substantially affected by spin-orbit interaction. In the CB, we find that the $K_C$ band lies below the $Q_C$ band. This conclusion is made on the basis of the magnitude of the conductance from respective bands (which arises in our computed results from the parallel wave-vectors of the bands), and is consistent with a recent study based on ARPES and detailed, first-principles computations [33]. However, in contrast, an STS study by Zhang et al. came to the conclusion that the $Q_C$ band actually lies below the $K_C$ band [34]. Their work employed measurements with variable tip-sample separations, as determined by constant-current operation (while scanning the sample voltage) in the STM, thereby achieving a measure of the voltage dependence of $\kappa$. They find an apparent drop in $\kappa$ right at the onset of the CB conductance spectrum compared to an energy ~0.1 eV or more higher up in the spectrum, from which they deduce that the parallel wave-vector of the lowest-most CB state is *less* than that of states located slightly higher in energy. While their data is of quite high quality, and their comparison with spectra acquired from a range of 2D materials is impressive, their result as to an apparent drop in $\kappa$ value right at the onset of the CB conductance nonetheless appears to be faulty. It is not easy to discern the reason behind this faulty conclusion. However, we note that the constant-current method for producing variable tip-sample separation can, in principle, lead to rather large changes in the separation, i.e., producing tip-sample separations that may end up being relatively small. For the case of a 2D heterobilayer, it was found in our prior work that these sorts of small tip-sample separation can lead to significant distortion of the STM images [10] (as also faintly seen in the data of Zhang et al. [16]), as the 2D layers are pressed together under the influence of the probe-tip. Such an effect, conceivably, might occur during the constant-current voltage scans utilized by Zhang et al. [34], thus possible giving rise to the apparent drop in the $\kappa$ value they observe at the onset of the CB conductance signal.

**Acknowledgements**

R.M.F. expresses his sincere gratitude to the many collaborators over the years, who made possible the development and application of the high-dynamic-range STS methods described in this work. R.M.F. was supported by the National Science Foundation (DMR-1809145). G.R.F. was supported by a Summer Undergraduate Research Fellowship from Carnegie Mellon University. Y.C.L., B.J., K.Z., and J.A.R. were supported by the Center for Atomically Thin Multifunctional Coatings (ATOMIC), sponsored by the National Science Foundation division of Industrial, Innovation & Partnership (IIP-1540018); and by the Semiconductor Research Corporation as the NEWLIMITS Center and NIST (70NANB17H041). Y.P. was supported by the National Key R&D Program of China (2017YFA0206202) and the National Science Foundation of China (11704303).

The data that support the findings of this study are available from the corresponding author upon reasonable request.

**References:**

[1] G. Binnig, H. Rohrer, Ch. Gerber, and E. Weibel, Phys. Rev. Lett. **49**, 57 (1983).

[2] R. S. Becker, J. A. Golovchenko, D. R. Hamann and B. S. Swartzentruber, Phys. Rev. Lett. **55**, 2012 (1985).

[3] R. M. Feenstra, W. A. Thompson and A. P. Fein. Phys. Rev. Lett. **56**, 608 (1986).

[4] R. J. Hamers, R. M. Tromp and J. E. Demuth, Phys. Rev. Lett. **56**, 1972 (1986).

[5] For review of early STS works, particularly those involving surface states, see R. M. Feenstra, Surf. Sci. **299/300**, 965 (1994).

[6] J. A. Stroscio, R. M. Feenstra and A. P. Fein. Phys. Rev. Lett. **57**, 2570 (1986).

[7] R. M. Feenstra and J. A. Stroscio, J. Vac. Sci. Technol. B **5**, 923 (1987).

[8] P. Mårtensson and R. M. Feenstra, Phys. Rev. B **39**, 7744 (1989).

[9] Y. Pan, S. Fölsch, Y. Nie, D. Waters, Y.-C. Lin, B. Jariwala, K. Zhang, K. Cho, J. A. Robinson, and R. M. Feenstra, Nano Lett. **18**, 1849 (2018).

[10] D. Waters, Y. Nie, F. Lüpke, Y. Pan, S. Fölsch, Y.-C. Lin, B. Jariwala, K. Zhang, C. Wang, H. Lv, K. Cho, D. Xiao, J. A. Robinson, and R. M. Feenstra, ACS Nano **14**, 7564 (2020).

[11] K. S. Novoselov, A. K. Geim, S. V. Morozov, D. Jiang, Y. Zhang, S. V. Dubonos, I. V. Grigorieva, and A. A. Firsov, Science **306**, 666 (2004).

[12] A. K. Geim and K. S. Novoselov, Nature Materials **6**, 183 (2007).

[13] A. Seabaugh, S. Fathipour, W. Li, H. Lu, J. H. Park, A. . Kummel, D. Jena, S. K. Fullerton-Shirey, and P. Fay, "Steep subthreshold swing tunnel FETs: GaN/InN/GaN and transition metal dichalcogenide channels," *2015 IEEE International Electron Devices Meeting (IEDM)*, Washington, DC, 2015, pp. 35.6.1-35.6.4, doi: 10.1109/IEDM.2015.7409835.

[14] C. G. Slough, B. Giambattista, A. Johnson, W. C. McNairy, C. Wang and R. V. Coleman, Phys. Rev. B **37**, 6571 (1988).

[15] R.E. Thomson, U. Walter, E. Ganz, J. Clarke, A. Zettl, P. Rauch, F.J. DiSalvo Phys. Rev. B **38**, 10734 (1988).

[16] M. Yankowitz, D. McKenzie, and B. J. LeRoy, Phys. Rev. Lett. **115**, 136803 (2015).

[17] J. H. Park, S. Vishwanath, X. Liu, H. Zhou, S. M. Eichfeld, S. K. Fullerton-Shirey. J. A. Robinson, R. M. Feenstra, J. Furdyna, D. Jena, H. G. Xing, and A. C. Kummel, ACS Nano **10**, 4258 (2016).

[18] C. Zhang, C.-P. Chuu, X. Ren, M.-Y. Li, L.-J. Li, C. Jin, M.-Y. Chou, C.-K. Shih, Sci. Adv. **3**, e1601459 (2017).

[19] R Addou, C. M. Smyth, J. Y. Noh, Y.-C. Lin, Y. Pan, S. M. Eichfeld, S. Fölsch, J. A. Robinson, K. Cho, R M Feenstra, and R. M. Wallace, 2D Mater. **5**, 025017 (2018).

[20] W. H. Blades, N. J. Frady, P. M. Litwin, S. J. McDonnell, and P. Reinke, J. Phys. Chem. C **124**, 1533 (2020).

[21] H. M. Hill, A. F. Rigosi, K. T. Rim, G. W. Flynn, and T. F. Heinz, Nano Lett. **16**, 4831 (2016).

[22] R. M. Feenstra, J. Vac. Sci. Technol. B **21**, 2080 (2003).

[23] Program SEMITIP, available at http://www.andrew.cmu.edu/user/feenstra/

[24] S. Gaan, G. He, R. M. Feenstra, J. Walker and E. Towe, J. Appl. Phys. **108**, 114315 (2010).

[25] R M Feenstra and S Gaan, J. Phys.: Conf. Ser. **326**, 012009 (2011).

[26] R. Dombrowski, Chr. Steinebach, Chr. Wittneven, M. Morgenstern, and R. Wiesendanger, Phys. Rev. B **59**, 8043 (1999).

[27] R. M. Feenstra, G. Meyer, F. Moresco, and K. H. Rieder, Phys. Rev. B **64**, 081306 (2001).

[28] N. Ishida, K. Sueoka, and R. M. Feenstra, Phys. Rev. B **80**, 075320 (2009).

[29] P. Ebert, L. Ivanova, and H. Eisele, Phys. Rev. B. **80**, 085316 (2009).

[30] R. M. Feenstra, Phys. Rev. B **50**, 4561 (1994).

[31] W. S. Yun, S. W. Han, S. C. Hong, I. G. Kim, and J. D. Lee, Phys. Rev. B **85**, 033305 (2012).

[32] D. Le, A. Barinov, E. Preciado, M. Isarraraz, I. Tanabe, T. Komescu, C. Troha, L. Bartels, T. S. Rahman, and P. A. Dowben, J. Phys.: Condens. Matter **27**, 182201 (2015).

[33] P. V. Nguyen, N. C. Teutsch, N. P. Wilson, J. Kahn, X. Xia, A. J. Graham, V. Kandyba, A. Giampietri, A. Barinov, G. C. Constantinescu, N. Yeung, N. D. M. Hine, X. Xu, D. H. Cobden, and N. R. Wilson, Nature **572**, 220 (2019).

[34] C. Zhang, Y. Chen, A. Johnson, M.-Y. Li, L.-J. Li, P. C. Mende, R. M. Feenstra, C.-K. Shih, Nano Lett. **15**, 6494 (2015).

[35] J. Tersoff and D. R. Hamann, Phys. Rev. Lett. 50, 1998 (1983).

[36] E. Stoll, A. Baratoff, A. Selloni, and P. Carnevali, J. Phys. C: Solid State Phys. **17**, 3073 (1984).

[37] A. Baratoff, Physica **127B**, 143 (1984).

[38] J. Tersoff and D. R. Hamann, Phys. Rev. B **31**, 805 (1985).

[39] Y. Pan, S. Fölsch, Y.-C. Lin, B. Jariwala, J. A. Robinson, Y. Nie, K. Cho, and R. M. Feenstra, 2D Mater. **6**, 021001 (2019).

[40] R. M. Feenstra, J. Y. Lee, M. H. Kang, G. Meyer, and K. H. Rieder, Phys. Rev. B **73**, 035310 (2006).

[41] R. M. Feenstra, Phys. Rev. Lett. **63**, 1412 (1989).

[42] R. M. Feenstra, G. Meyer and K. H. Rieder, Phys. Rev. B **69**, 081309 (2004).

[43] C. B. Duke, *Tunneling in Solids* (Academic Press, New York, 1969).

[44] J. G. Simmons, J. Appl. Phys. **34**, 1793 (1963).

[45] J. Bardeen, Phys. Rev. Lett. **6**, 57 (1961).

[46] R. M. Feenstra, D. Jena, and G. Gu, J. Appl. Phys. **111**, 043711 (2012).

[47] C. Kittel, *Introduction to Solid State Physics*, 8th edition (John Wiley and Sons, New Jersey, 2005).

[48] G. Frazier, Y. Pan, S. Fölsch, Y.-C. Lin, B. Jariwala, K. Zhang, J. A. Robinson, and R. M. Feenstra, to be published.

[49] Y. Dong, R. M. Feenstra, M. P. Semtsiv, and W. T. Masselink, J. Appl. Phys. **103**, 073704 (2008).

[50] Supplementary Material at URL xxx.

[51] S. C. de la Barrera, Q. Gao, and R. M. Feenstra, J. Vac. Sci. Technol. B **32**, 04E101 (2014).

**Acquisition and Analysis of Scanning Tunneling Spectroscopy Data – WSe₂ Monolayer**

Randall M. Feenstra,[1] Grayson R. Frazier,[1] Yi Pan,[2,3] Stefan Fölsch,[2] Yu-Chuan Lin,[4] Bhakti Jariwala,[4] Kehao Zhang,[4] and Joshua A. Robinson[4]

[1]Dept. Physics, Carnegie Mellon University, Pittsburgh, PA, 15213 U.S.A.

[2]Paul-Drude-Institut für Festkörperelektronik, Hausvogteiplatz 5-7, 10117 Berlin, Germany

[3]Center for Spintronics and Quantum Systems, State Key Laboratory for Mechanical Behavior of Materials, Xi'an Jiaotong University, Xi'an 710049, China

[4]Dept. Materials Science and Engineering, and Center for 2-Dimensional and Layered Materials, The Pennsylvania State University, University Park, PA, 16802 U.S.A.

**SUPPLEMENTAL MATERIAL**

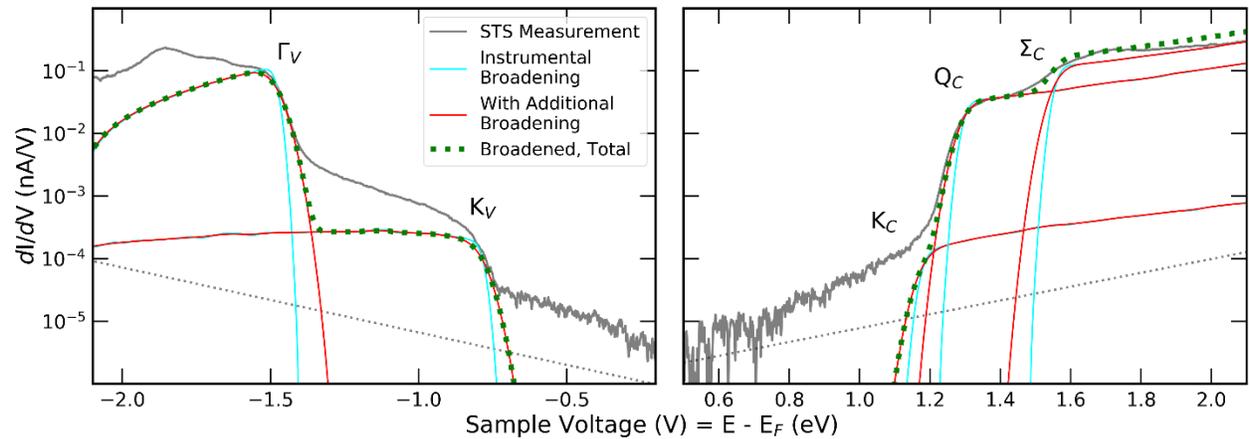

FIG S1.  Fits to WSe₂ monolayer spectra, using parabolic dispersion for the assumed bands rather than the hyperbolic dispersion in the main text (the same effective mass and onset energies are used for both parabolic and hyperbolic bands, and the same broadening parameters are applied in both cases). The main difference between the parabolic and hyperbolic dispersion is the shape of the computed bands *beyond* their onsets (i.e. above the onset for a CB or below for a VB), which are seen to be significantly flatter in these plots compared to the main text.

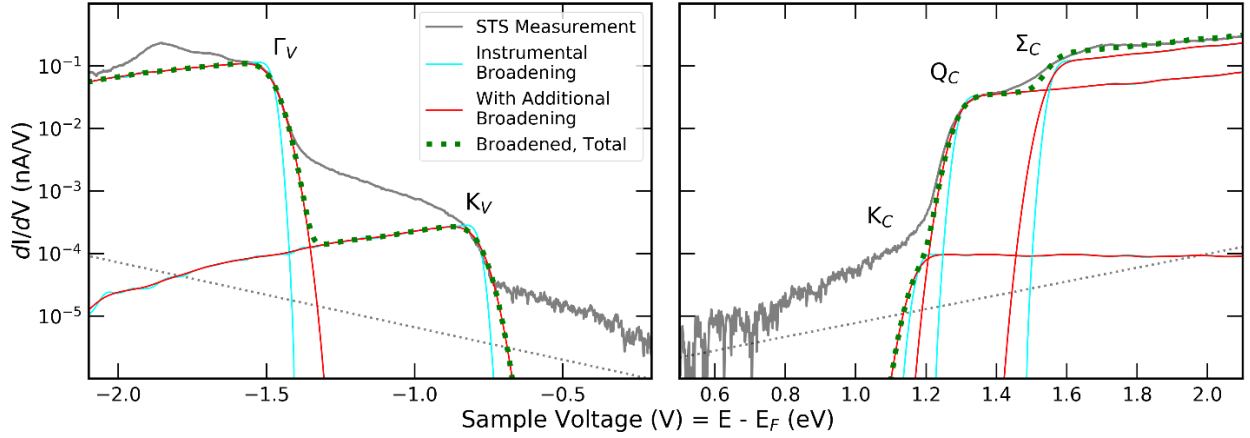

FIG S2. Fits to WSe$_2$ monolayer spectra, using parabolic dispersion for the assumed bands and also neglecting the variation in parallel wave-vector that occurs in the bands (i.e. using $k_{||}^{(0)} \equiv \left| \mathbf{k}_{||}^{(0)} \right|$ rather than $k_{||} \equiv |\mathbf{k}_{||}|$ in the transmission term, Eq. (15) of the main text). The effect of this neglect of the variation in parallel wave-vector is seen to be a downturn (or less rapid increase) in the conductance beyond the onsets of the bands, compared to the results of Fig. S1.

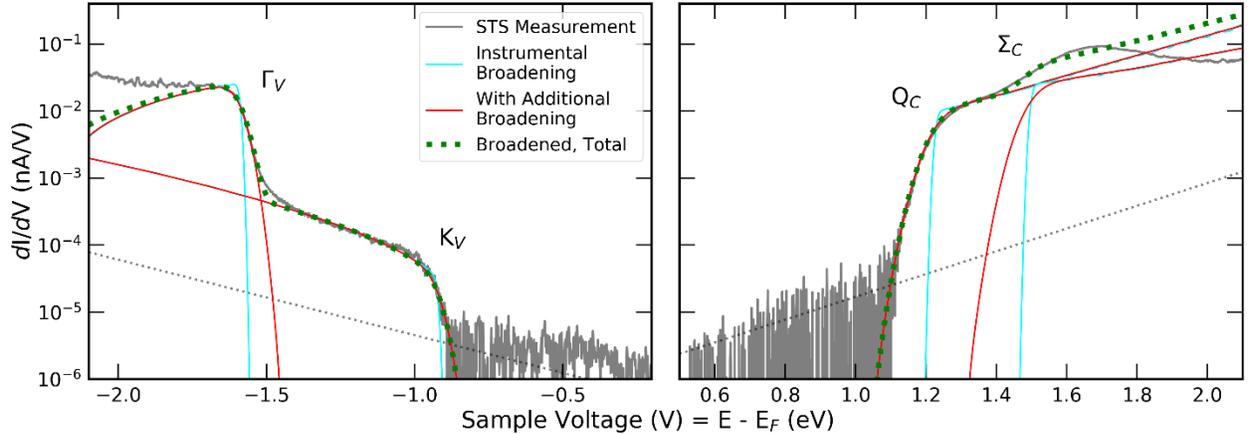

FIG S3. STS data and curve fitting for a spectrum acquired at a different spatial location on the surface (and using a different probe tip) than the data of the main text. The same effective masses and band velocities as listed in Table I of the main text were used for the fitting, although small changes in the band onsets compared to Table I were utilized (shifts of $-0.11$ V for the $\Gamma_V$ band, $-0.15$ V for the K$_V$ band, and $-0.06$ V for the Q$_C$ and $\Sigma_C$ bands). The rms modulation voltage of the lock-in amplifier used for acquisition of this data was 7.1 mV, i.e., 2.5$\times$ smaller than for the data in the main text. Additional broadening amounts of 33.9 and 43.9 mV were utilized for the VB and CB fits, respectively. No attempt at EG background subtraction for the CB conductance has been made (since the signal-to-noise of this data is significantly worse than for Fig. 4), so no K$_C$ band is fit to the data.